\documentclass[prl,twocolumn,10pt,showpacs]{revtex4-1}

\usepackage[TS1,OT1,T1]{fontenc}
\usepackage{hyperref}
\usepackage{graphicx}
\usepackage{sidecap}
\usepackage{verbatim}
\usepackage{bm}
\usepackage{amssymb}

\begin{document}
\title{Spin-orbit coupling and spin Hall effect for neutral atoms without spin-flips}
\author{Colin J. Kennedy, Georgios A. Siviloglou, Hirokazu Miyake, William Cody Burton, and Wolfgang Ketterle}
\affiliation{MIT-Harvard Center for Ultracold Atoms, Research Laboratory of Electronics, Department of Physics, Massachusetts Institute of Technology, Cambridge, Massachusetts 02139, USA}
\date{\today}
\begin{abstract}
We propose a scheme which realizes spin-orbit coupling and the spin Hall effect for neutral atoms in optical lattices without relying on near resonant laser light to couple different spin states.  The spin-orbit coupling is created by modifying the motion of atoms in a spin-dependent way by laser recoil. The spin selectivity is provided by Zeeman shifts created with a magnetic field gradient.  Alternatively, a quantum spin Hamiltonian can be created by all-optical means using a period-tripling, spin-dependent superlattice.
\end{abstract}
\maketitle
Many recent advances in condensed matter physics are related to the spin degree of freedom.  The field of spintronics \cite{spin12}, the spin Hall effect \cite{kato04}, and topological insulators \cite{hasa10} all rely on the interplay between spin and motional degrees of freedom provided by spin-orbit coupling.  Quantum simulations with neutral atoms have started to implement spin-orbit coupling using Raman transitions between different hyperfine states \cite{lin11,zhan12,wang13,cheu13,gali13}.  Since the Raman process transfers momentum to the atom,  the resonance frequency is Doppler sensitive, and thus couples motion and spin.

The possibility of using spin-flip Raman processes to create interesting gauge fields was first pointed out in \cite{higb02,juze04,juze06}, and extended to non-Abelian gauge fields, which imply spin-orbit coupling, in \cite{ruse05,oste05}. With the exception of an atom chip proposal where the spin-flips are induced with localized microwave fields \cite{ande13}, all recently proposed schemes are based on spin-flip Raman processes \cite{juze10,dali11,ho11,ande12,xu12,cole12,gali13}.

The major limitation of these Raman schemes is that spin-flip processes are inevitably connected with heating by spontaneous emission if they rely on spin-orbit coupling in the excited state, as in alkali atoms or other atoms with an $S$ orbital ground state.  Since laser beams interact with atoms via the electric dipole interaction, they do not flip the spin.  Spin-flips occur only due to intrinsic spin-orbit interactions within the atoms; therefore, spin-orbit coupling by spin flip Raman processes relies on the spin-orbit coupling \emph{within} the atom. Since the spontaneous emission rate and the two-photon Rabi frequency for Raman spin-flip processes scale in the same way with respect to the ratio of laser power to detuning, for a given atom the coupling strength relative to the spontaneous emission rate is fixed by the fine structure splitting compared to the natural linewidth. This has not been a limitation for the demonstration of single-particle or mean-field physics \cite{lin11,zhan12,wang13,cheu13,gali13}, but will become a severe restriction for many-body physics where the interactions will introduce a smaller energy scale and therefore require longer lifetimes of the atomic sample. Some authors have considered transitions involving metastable states of alkaline earth atoms to reduce the effects of spontaneous emission \cite{gerb10,beri11}.

Here we present a spin-orbit coupling scheme that does not involve spin-flips, is diagonal in the spin component, $\sigma_z$, and corresponds to an Abelian $SU(2)$ gauge field. This scheme can be implemented with far-off resonant laser beams, thus overcoming the limitation of short sample lifetimes due to spontaneous emission. In the field of cold atoms, many discussions of spin-orbit coupling emphasize its close relationship to non-Abelian gauge fields \cite{ho11,beel13} which are non-diagonal for any spin component and therefore mix spin and motion in a more complicated way. However, a scheme diagonal in the spin component is sufficient for spin Hall physics and topological insulators \cite{kane05a,bern06}, and its implementation has major experimental advantages. In the theoretical proposals \cite{liu07,zhu06} and the demonstration \cite{beel13} of the spin Hall effect for quantum gases, Raman spin-flips are used to create an Abelian gauge field diagonal with respect to one spin component.

The physical principle of the spin-orbit coupling scheme presented here is very different from spin-flip schemes. It does not require any kind of spin-orbit coupling within the atom. Rather, spin-dependent vector potentials are engineered utilizing the Zeeman effect in a magnetic field -- atoms in the spin up and down states interact with different pairs of laser beams, or differently with the same pair, and the photon recoil changes the atom's motion in a spin-dependent way. This results in spin-orbit coupling which is diagonal in the spin basis.

To begin, we summarize the relationship between spin-orbit coupling and spin-dependent vector potentials. For charged particles, the origin of spin-orbit coupling is the relativistic transformation of electromagnetic fields.  When an electron moves through an electric field $\mathbf{E}$, it experiences a magnetic field $\mathbf{B}$ in its moving frame which interacts with the spin $\bm{\sigma}$ (described by  the Pauli spin matricies).  Spin-orbit coupling contributes a term proportional to $(\mathbf{p} \times \mathbf{E})\cdot \bm{\sigma}$ in the Hamiltonian. As such, an electric field in the $z$-direction gives rise to the Rashba spin-orbit coupling $(\bm{\sigma} \times \mathbf{p})_z=\sigma_x p_y-\sigma_y p_x$.

Assuming a 2D system confined to the $x, y$ plane, and an in-plane electric field,  the spin-orbit interaction conserves $\sigma_z$.  Following \cite{bern06}, a radial electric field $\mathbf{E} \sim E(x,y,0)$ leads to a spin-orbit coupling term in the Hamiltonian of the form $E \sigma_z (xp_y-yp_x)$.  Such a radial field could be created by a uniformly charged cylinder, or can be induced by applying stress to a semiconductor sample \cite{bern06}.  This spin-coupling term is identical to the $\mathbf{A}\cdot \mathbf{p}$ term for the Hamiltonian describing a spin in a magnetic field, $\sigma_z B$. Using the symmetric gauge for the vector potential $\mathbf{A} = \frac{\sigma_z B}{2}(y, -x,0)$, one obtains a term in the Hamiltonian proportional to $\sigma_z B(xp_y-yp_x)$ or equivalently to $B\bm{\sigma}\cdot \mathbf{L}$, where $\mathbf{L}$ is the orbital angular momentum of the atom.  Therefore, this form of spin-orbit coupling is equivalent to a spin-dependent magnetic field which exerts opposite Lorentz forces on spin up and down atoms. This leads to the spin Hall effect which creates a transverse spin current and no charge or mass currents \cite{kane05a, bern06}. The $\mathbf{A}^2$ term constitutes a parabolic spin-independent potential which is irrelevant for the spin physics discussed here.

We now present a scheme which realizes such an Abelian gauge field and manifests itself as a spin-dependent magnetic field.  Recently, the MIT group \cite{miya13,miya13th} and the Munich group \cite{aide13,aide13b} have suggested  and implemented a scheme to generate synthetic magnetic fields for neutral atoms in an optical lattice.  The scheme is based on  the simple Hamiltonian for non-interacting particles in a 2D cubic lattice,
\begin{equation} \label{H_vanilla}
H=-\sum_{m,n} \big(J_x \hat a^{\dagger}_{m+1,n}\hat a_{m,n} + J_y\hat a^{\dagger}_{m,n+1}\hat a_{m,n} + h.c. \big)
\end{equation}
where $J_{x(y)}$ describes tunneling in the  $x$-($y$-)direction and $\hat a^{\dagger}_{m,n}$ ($\hat a_{m,n}$) is the creation (annihilation) operator of a particle at lattice site $(m,n)$. The setup is detailed in \cite{miya13} and summarized as follows: a linear tilt of energy $\Delta$ per lattice site is applied using a magnetic field gradient in the $x$-direction, thus suppressing normal tunneling in this direction. Resonant tunneling is restored with two far-detuned Raman beams of two-photon Rabi frequency $\Omega$, frequency detuning $\delta \omega = \omega_1 - \omega_2$, and momentum transfer $\mathbf{k} = \mathbf{k_1} - \mathbf{k_2}$. Considering only the case of resonant tunneling, $\delta \omega = \Delta/\hbar$, rapidly oscillating terms time average out \cite{jaks03}, yielding an effective Hamiltonian which is time-independent \cite{miya13}.

\begin{equation} \label{Harper}
H=-\sum_{m,n} \big( Ke^{-i\phi_{m,n}}\hat a^{\dagger}_{m+1,n}\hat a_{m,n} + J\hat a^{\dagger}_{m,n+1}\hat a_{m,n} +h.c. \big)
\end{equation}

\begin{figure}
\centering
\includegraphics[width=0.48\textwidth]{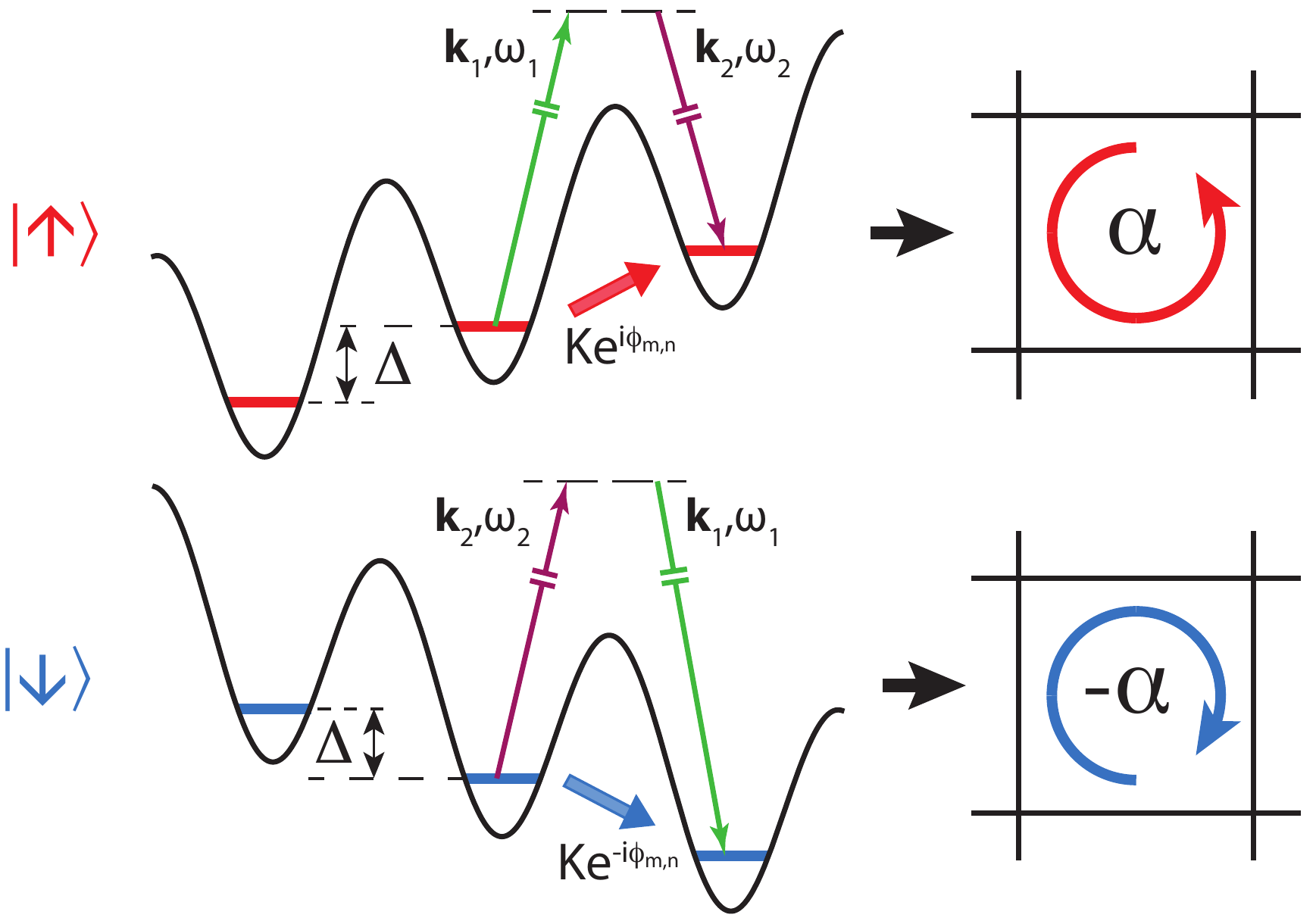}
\caption{Spin-dependent tunneling in an optical lattice tilted by a magnetic field gradient.  When the two spin states have opposite magnetic moments, then the role of absorbtion and emission of the two photons is exchanged.  The result is that the two states have tunneling matrix elements with opposite phases, leading to opposite synthetic magnetic fields and realizing spin-orbit coupling and the quantum spin Hall effect.}
\end{figure}

This effective Hamiltonian describes charged particles on a lattice in a magnetic field under the tight-binding approximation \cite{harp55,hofs76}.
The gauge field arises from the spatially-varying phase $\phi_{m,n}  =  \mathbf{k}\cdot\mathbf{R}_{m,n}= m k_x a + n k_y a$  where $a$ is the lattice constant and has the form $\mathbf{A} = \hbar (k_x x + k_y y)/a\: \hat{\mathbf{x}}$. One can tune the flux per unit cell, $\alpha$, for a given spin state over the full range between zero and one by adjusting the angle between the Raman beams, and consequently $k_y$.

We now extend this scheme to the spin degree of freedom, and assume a mixture of atoms in two hyperfine states, labeled spin up and down.  If the potential energy gradient is the same for the two states, then the two states experience the same magnetic field.  This is the situation when the tilt is provided by gravity, a scalar AC Stark shift gradient, or a magnetic field gradient if both states have the same magnetic moment -- the phase $\phi_{m,n}$ is independent of $\sigma_z$.

If the two states have the same value of the magnetic moment, but opposite sign, then the potential gradient is opposite for the two states. This can be realized by using states of the same hyperfine level $F$, but with opposite magnetic quantum number $M_F$ (e.g. in $^{23}$Na or $^{87}$Rb, the $|F,M_F\rangle=|2,2\rangle$ and $|2,-2\rangle$ states), or by picking another suitable pair of hyperfine states.  In this case, for laser-assisted tunneling between two sites $m$ and $m+1$, the roles of the two laser beams -- absorption of a photon versus stimulated emission of a photon -- for the Raman process are reversed as depicted in Fig. 1.  Therefore, the two states receive opposite momentum transfer, and this sign change leads to a sign change for the enclosed phase:
\begin{equation}
\phi_{m,n}=(m k_x a + n k_y a)\:\sigma_z
\end{equation}
and also for the vector potential and the magnetic field.  The vector potential realized by this scheme:
\begin{equation}
\mathbf{A} = \frac{\hbar}{a} (k_x x + k_y y)\: \hat{\mathbf{x}}\:\hat{\sigma}_z
\end{equation}
creates the spin-orbit coupling  discussed in the introduction, although in a different gauge. The $x$-dependence in the $x$-component of $\mathbf{A}$ is necessary for a non-negligible tunneling matrix element for the laser-assisted process \cite{miya13}.

This system has now time reversal symmetry, in contrast to the system with the same synthetic magnetic field for both states (since a magnetic field breaks time reversal symmetry).  It therefore realizes the quantized spin Hall effect consisting of two opposite quantum Hall phases. It is protected by a $\mathbb{Z}$ topological index due to fact that $\sigma_z$ is conserved \cite{kane05a, bern06}.

When the values of the two magnetic moments are different, and the potential energy gradient is provided by a magnetic field gradient, then the two states have different Bloch oscillation frequencies, $\Delta/h$.  Each state now needs two separate Raman beams for laser-assisted tunneling (or they can share one beam).  This implies that the synthetic magnetic field can now be chosen to be the same, to be opposite or to be different for the two spin states.  One option is to have zero synthetic magnetic field for one of the states.  Atoms in this state can still tunnel along the tilt direction by using a Raman process without $y$-momentum transfer, or equivalently, by inducing tunneling through lattice modulation \cite{ivan08}. In the case of two different magnetic moments, one could also perform dynamic experiments, where laser parameters are modified in such a way that one switches either suddenly or adiabatically from the quantum Hall effect to the spin quantum Hall effect.

An intriguing possibility is to couple the two states.  Since $\sigma_z$ is no longer conserved, the system should become a topological insulator with the $\mathbb{Z}_2$ classification \cite{kane05b,hasa10}, provided that the coupling is done in a time-reversal invariant way.  This can be done with a term which is not diagonal in $\sigma_z$ -- ie. a  $\sigma_x p_y$ term -- by adding spin-flip Raman lasers to induce spin-orbit coupling, or by driving the spin-flip transition with RF or microwave fields.  A coherent RF drive field would not be time-reversal invariant, but it would be interesting to study the effect of symmetry-breaking in such a state \cite{gold12}.  A drive field where the phase is randomized should lead to a time-reversal invariant Hamiltonian.

Our scheme implements the idealized scheme for a quantum spin Hall system consisting of two opposite quantum Hall phases.  This is a starting point for breaking symmetries and exploring additional terms in the Hamiltonian.  Ref. \cite{gold12, beug12} discusses a weak quantum spin Hall phase, induced by breaking the time-reversal symmetry by a magnetic field - this can be achieved by population imbalance between the two spin states.  A spin-imbalanced quantum Hall phase can turn into a spin-filtered quantum Hall phase \cite{gold12, beug12} where only one component has chiral edge states.  This can be achieved by realizing a finite synthetic magnetic field for one component, and zero for the other.  Changing the spin-orbit coupling can induce topological quantum phase transitions between a helical quantum spin Hall phase and a chiral spin-imbalanced quantum Hall state.  This can probably be achieved in a population imbalanced system by adding additional Raman spin-flip beams \cite{gold12, beug12}.

So far, we have discussed single-particle physics.  Adding interactions, by increasing the density with deeper lattices or through Feshbach resonances, will induce interesting correlations and may lead to fractional topological insulators \cite{levi09}.  Another option are spin-drag experiments \cite{duin09, koll12}, transport experiments where one spin component transfers momentum to the other component.  For the situation mentioned above, where the synthetic magnetic field is zero for one component (e.g. spin up), a transport experiment revealing the Hall effect \cite{lebl12} for spin down would show a non-vanishing Hall conductivity for spin up to due to spin drag. In addition, one would expect that spin-exchange interactions destroy the two opposite quantum Hall phases, and should lead to the quantum spin Hall phase with $\mathbb{Z}_2$ topological index.

We now present another way of realizing the physics discussed above, using optical superlattices instead of a potential energy gradient.  This has the advantage of purely optical control, and avoids possible heating due to Landau-Zener tunneling \cite{niu96} between Wannier Stark states.  So far, optical superlattices have allowed the observation of the ground state with staggered magnetic flux \cite{aide11}, in contrast to experiments with magnetic tilts \cite{miya13,aide13b}.

\begin{figure}
\centering
\includegraphics[width=0.48\textwidth]{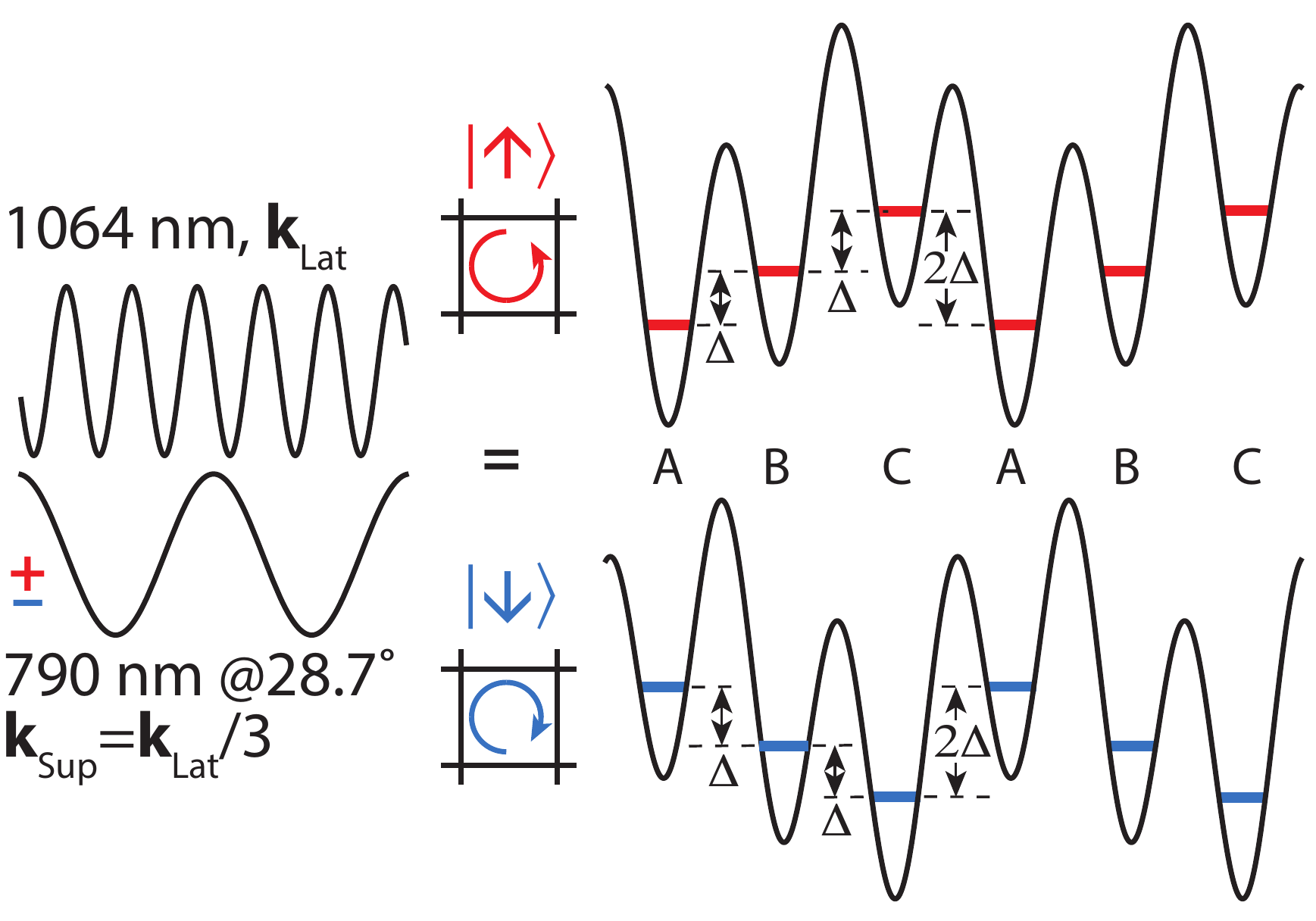}
\caption{Superlattice scheme for realizing the quantnum Hall and quantum spin Hall effect. A superlattice with three times the spatial period as the fundamental lattice leads to three distinguishable sites A, B, C.  For the quantum spin Hall effect, the superlattice operates at a magic wavelength where the AC Stark effect is opposite for the two spin states.  For rubidium, this is achieved at a wavelength of 790 nm.}
\end{figure}

Figure 2 summarizes the new scheme.  The superlattice has three times the period of the basic lattice, thus distinguishing sites A, B, C in energy.  Resonant tunneling is re-established using three pairs of Raman beams with frequencies: $\omega_1 + \Delta_{AB}/\hbar$, $\omega_2 + \Delta_{BC}/\hbar$, and $\omega_3 - (\Delta_{AB}+\Delta_{BC})/2\hbar$ collinear in one arm and $\omega_1$, $\omega_2$, and $\omega_3$ collinear in another arm at an angle to the first. Consequently, there is always the same momentum transfer for tunneling in the $y$-direction, leading to the same flux as the scheme with the magnetic tilt, and Eqs. (3) and (4)  apply.  This is in contrast to schemes with two distinguished sites A and B (by using internal states \cite{jaks03,gerb10} or a superlattice \cite{aide11}) which lead to a staggered magnetic field.  Rectification of the magnetic flux in a staggered configuration by adding a tilt \cite{jaks03, aide11} or a superlattice \cite{gerb10} has also been proposed.  In the latter scheme, this would result in four distinguishable sites (two internal states A, B, doubled up by the superlattice).  Another rectification scheme uses three internal states \cite{muel04}. Our scheme avoids spin-flip transitions between internal states, and has the minimum number of ingredients of three different sites to provide directionality.  Furthermore, by adjusting the spatial phase shift between the fundamental and the superlattice, one can choose the energy offsets $\Delta_{AB}=\Delta_{BC}=-\Delta_{CA}/2$ (see Figure 2).  The scheme can then be implemented by shining Raman beams from two directions, each beam having two frequencies.

This scheme would realize Hofstadter's butterfly and the quantum Hall effect.  For the quantum spin Hall effect, one has to choose the superlattice laser to be at the magic wavelength where the scalar AC Stark shift vanishes, and only a vector AC Stark shift remains corresponding to a so-called fictitious magnetic field \cite{mcka10,cohe72}. By detuning the laser between the $D_1$ and $D_2$ lines, one can achieve a pure vector AC Stark shift, which is equal in magnitude, but opposite in sign when the atoms in the two hypefine states have opposite magnetic moments.  In this case, the superlattice will provide opposite potentials for the two states, resulting in opposite momentum transfers due to the Raman beams and opposite vector potentials. The superlattice period is $\lambda/(2 \sin{(\theta/2}))$, where $\theta$ is the angle between the two superlattice beams, which is adjusted to make the superlattice period three times the period of the basic lattice.  This scheme realizes the quantum spin Hall effect and a topological insulator with two opposite quantum Hall phases with a purely optical scheme and no Raman spin-flip transitions.

To replace the magnetic field gradient by a superlattice that generates a fictitious magnetic field, the laser detuning has to be on the order of the fine structure splitting, resulting in heating due to spontaneous emission.  For atoms like rubidium, the lifetime is many seconds \cite{mcka10}. To be specific, we consider a low-density gas Rb atoms in the $F = 2 $, $M_F = \pm 2$ states in a lattice with a depth of ten photon recoils at the wavelength of 1064 nm. A superlattice with a lattice depth of 10 kHz is created by interfering two laser beams at 790.0 nm of 1.0 mW of laser power and a beam waist of 125 $\mu$m. The resulting offset $\Delta_{AB}$ and $\Delta_{AC}$ will be approximately 4 and 8 kHz respectively, well placed in the bandgap of the basic lattice. The spontaneous scattering rate induced by the superlattice beams is less than 0.1$/$ s. Alternatively, the superlattice producing the fictitious magnetic field can be replaced by a sinusoidal (real) magnetic field generated by an atom chip \cite{gold10}.

There have been several suggestions how to detect properties of the quantum Hall and  quantum spin Hall phases.  Time-of-flight pictures will reveal the enlarged magnetic unit cell due to the synthetic magnetic field \cite{moel10, powe11, aide11, pola13}.  Hall plateaus can be discerned in the density distribuion \cite{umcu08}.  The Chern number of a filled band can be measured interferometically \cite{aban13} or using ballistic expansion \cite{zhao11}.  Topological edge states can be directly imaged \cite{stan09, gold13} or detected by Bragg spectroscopy \cite{liu10, gold12a, buch12}.

Our work maps out a route towards spin-orbit coupling, the spin Hall effect and topological insulators which does not require coupling of different internal states with spin-flipping Raman lasers.  The Hamiltonian describing the system is diagonal in the $\sigma_z$ spin component.  This follows closely the two original papers on the spin Hall effect \cite{kane05a,bern06}.  In addition, we have presented two configurations for realizing a quantum spin Hall Hamiltonian.  The scheme with the magnetic tilt completely avoids near resonant light, and the superlattice scheme provides a  purely optical approach.

This work was supported by the NSF through the Center of Ultracold Atoms, by NSF award PHY-0969731, under ARO Grant No. W911NF-13-1-0031 with funds from the DARPA OLE program, and by ONR.  This work was completed at the Aspen Center for Physics (supported in part by the National Science Foundation under Grant No. PHYS-1066293), and insightful discussions with  Hui Zhai, Jason Ho, and Nigel Cooper are acknowledged. We thank Wujie Huang for a critical reading of the manuscript.

After most of this work was completed \cite{kett13talk} we became aware of similar work carried out in the group of I. Bloch in Munich \cite{bloc13talk,aide13b}.

\end{document}